\documentclass[prd,aps,nofootinbib,superscriptaddress,preprint]{revtex4-1}
\usepackage{amsmath,amssymb,amsfonts,color,graphicx,graphics,latexsym,placeins,epsfig,multirow}
\usepackage{array}
\newcolumntype{P}[1]{>{\centering\arraybackslash}p{#1}}
\usepackage{footmisc}
\usepackage[compatibility=false]{caption}
\usepackage{subcaption}
\usepackage{comment}
\usepackage{enumitem}
\setdescription{leftmargin=8pt}
\usepackage{hyperref}
\hypersetup{
	colorlinks=true,       % false: boxed links; true: colored links
	linkcolor=blue,          % color of internal links
	citecolor=blue,        % color of links to bibliography
}

\usepackage{natbib}
\setcitestyle{square, comma, numbers, sort&compress}
\usepackage{float}

\usepackage{amsmath,amssymb}
\newcommand{\ba}{\begin{align}}
\newcommand{\ea}{\end{align}}

\def\3nab{\tilde{\nabla}}

\def\be {\begin{equation}}
\def\ee {\end{equation}}
\def\ba {\begin{eqnarray}}
\def\ea {\end{eqnarray}}

\newcommand{\sfr}[2]
{{\textstyle\frac{#1}{#2}}}

\newcommand{\df}[2]{{\dfrac{#1}{#2}}}

\newcommand{\barray}{\begin{array}}
\newcommand{\earray}{\end{array}}

\newcommand{\bea}{\begin{eqnarray}}
\newcommand{\eea}{\end{eqnarray}}

\usepackage[dvipsnames]{xcolor}
\begin{document}
%%%%%%%%%%%%%%% AUTHOR'S NAMES AND AFFILIATIONS %%%%%%%%%%%%%%%%%%
\title{\Large \bf Gravitational radiation from binary systems in Unimodular gravity}

\author{Indranil Chakraborty}
\email{indranil.phy@iitb.ac.in}
\affiliation{Department of Physics,  Indian Institute of Technology Bombay, Mumbai 400076, India}

\author{Soumya Jana}
\email{soumyajana.physics@gmail.com (Corresponding Author)}
\affiliation{Department of Physics, Sitananda College, Nandigram, 721631, India}

\author{Subhendra Mohanty}
\email{mohantys@iitk.ac.in}

\affiliation{Department of Physics, Indian Institute of Technology Kanpur, Kalyanpur, Kanpur 208016, India}
%%%%%%%%%%%%%%%%%%%%%%%%%%% ABSTRACT  %%%%%%%%%%%%%%%%%%%%%%%%%%

\begin{abstract}
Unimodular gravity (UG) is classically considered identical to General Relativity (GR). However, due to restricted diffeomorphism symmetry, the Bianchi identites do not lead to the conservation of energy-momentum tensor. Thus, the conservation of energy-momentum tensor needs to be separately assumed in order to reconcile with GR. Relaxing this assumption, one finds that the conservation violation can lead to differences with GR, which can be subsequently examined in astrophysical and cosmological scenarios. To this end, we examine the predictions of UG in the context of binary systems emitting gravitational radiation. Primarily, we show how the field equations involve a diffusion function which quantifies the measure of non-conservation. Due to this violation, the dispersion relation is modified. Incorporating these changes, we provide an expression for the energy loss by the binaries, which reduces to Peters-Mathews result in the GR limit. Using binary pulsar data, we constrain the theory parameter $\zeta$ (which signifies non-conservation) by determining the rate of orbital decay. The strongest constrain on $\zeta$ comes out to be $\vert \zeta \vert \leq 5\times 10^{-4}$ which is better by an order of magnitude than an existing equivalent constraint coming from the tidal deformability of the neutron stars.

\end{abstract}

%%%%%%%%%%%%%%%%%%%%%%%%%%%%%%%%%%%%%%%%%%%%%%%%%%%%%%
\maketitle
%%%%%%%%%%%%%%%%%%%%%%%%%%% INTRODUCTION  %%%%%%%%%%%

\section{introduction}

%thoughts---i) briad how GR is used, ii) how issues with cosmology cause problems, iii) how UG is instrumental in solving them, iv) mention some issues

General Relativity (GR) is widely regarded as the most successful classical theory of gravity since its inception in 1915. However, shortly following the introduction of GR, numerous other alternative theories of gravity have been put forward. Unimodular gravity (UG) has long been considered a viable alternative to GR. For a detailed discussion on this alternative formulation of a theory of gravity, see the review \cite{RevModPhys.61.1}.
UG have similarities with GR in several aspects. One significant similarity between the two theories is that they are both geometric theories of gravity that rely on a Lagrangian which includes the Ricci scalar. The novel addition, referred to as UG, involves maintaining the determinant of the metric of spacetime as a constant rather than allowing it to vary as a dynamical variable. This restriction limits the symmetry of the diffeomorphism group to the group of unimodular general coordinate transformations, usually referred to as the Transverse Diffeomorphism (see  \cite{Lopez-Villarejo:2010uib} for a comprehensive review). As a result, the field equations in UG are the trace-free Einstein field equations. Therefore, the vacuum energy in UG does not have a direct gravitational impact, and the cosmological constant is merely an integration constant of the dynamics. Weinberg noted that UG may be employed to address the cosmological constant problem, based on this characteristic \cite{RevModPhys.61.1}.

The Bianchi identities in the context of GR theory result in the automatic conservation of the Energy-Momentum (EM) tensor due to the general diffeomorphism invariance. The invariance of UG under a restricted class of transformations, as opposed to general diffeomorphisms, has a specific implication for the divergence of the EM tensor. It can either be zero, indicating the presence of the usual conservation law and the recovery of the GR equations with an additional integration constant that corresponds to the cosmological constant, or the EM tensor does not conserve, leading to the emergence of a completely new structure. Our present article aims to examine the second option. The investigation of UG, with or without the concept of conservation of the EM tensor \cite{10.1063/1.530131,Finkelstein:2000pg,Perez:2018wlo,Agrawal:2022atn,Nakayama:2022qbs,Almeida:2022qld}, has been thoroughly explored in the literature, particularly in the fields of cosmology \cite{Shaposhnikov:2008xb,Jain:2012gc,Garcia-Aspeitia:2019yni,Leon:2022kwn,Fabris:2021atr,LINARESCEDENO2021100807,Linares2023} and the quantisation problem of gravity. In \cite{PhysRevD.102.023508}, the authors used cosmological data to constrain the EM tensor non-conservation due to the restricted symmetry in UG. The analysis of light bending and Mercury's orbit with a given diffusion model in UG gravity was done in \cite{Bonder:2022kdw}. A recent study examines the evolution of gravitational waves in a universe governed by UG without conservation of the EM tensor. The study highlights the differences between the observed signatures of these waves and the standard signatures predicted by GR \cite{sym14010087}. Tidal deformability of neutron stars was studied in the context of UG and non-conservation of the EM tensor was tested in \cite{Yang:2022ees}. Very recently, the initial value problem in UG has been investigated in \cite{Herrera:2024sae}.

In the present article, we study UG without the conservation of the EM tensor in the context of the quasi-stationary binary systems, such as, binary neutron stars or  neutron star- white dwarf systems. Specifically, we derive the rate of energy loss of such binary systems due to gravitational radiation using a one-graviton vertex process and compare the predictions with GR and also constrain the non-conservation parameter in UG from the observations of the orbital period decay of the binary star systems. GR can be approximated as a quantum field theory of spin-2 fields in the Minkowski space when considering weak gravitational fields \cite{Feynman:1996kb, Weinberg:1964ew,Veltman:1975vx, Donoghue:2017pgk}. The interaction between massive bodies, such as the Newtonian potential or the bending of light by a heavy body, can be accurately described by the exchange of gravitons at the tree level. The outcome of the tree level exchange should correspond to the weak field limit of the classical GR results. The process of gravitational radiation arising from binary stars, represented as a single vertex Feynman diagram involving the emission of massless gravitons from a classical source, has been analysed in previous studies \cite{Mohanty:1994yi,Mohanty:2020pfa}. The findings of these studies are consistent with the conclusions obtained by Peter and Mathews~\cite{Peters:1963ux}, who employed the quadrupole formula of GR. This technique was also employed in the calculation of other types of radiation, such as vector gauge boson radiation \cite{KumarPoddar:2019ceq}, massive graviton radiation \cite{Poddar2022}, and radiation in $f(R)$ gravity \cite{Narang_2023}. We organize our article as follows. In Section~\ref{sec:formalism}, the theoretical formalism of UG without the conservation of the EM tensor is discussed. Next, in the Section~\ref{sec:radiation} we derive the gravitational radiation from binary star-system starting from the effective linearized UG action and following the method described in \cite{Mohanty:2020pfa,Poddar2022}. Using the obtained formula of the rate of energy loss due to graviton radiation we compare the theoretical prediction of orbital period decay with the observations in the Section~\ref{sec:obs} and obtain constrain the non-conservation parameter of UG. Finally, in the Section~\ref{sec:conclusions}, we summarize and discuss our results.

Throughout the paper we have used the natural system of units: $\hbar=c=1$, and $8\pi G=1/{M^2_{pl}}$ where Planck mass $M_{pl}=2.435\times 10^{18}$ GeV.

\section{Formalism}
\label{sec:formalism}

\subsection{Action functional in UG}

\noindent  UG action has been studied in different {\em avatars} yielding equivalent results \cite{Anderson:1971,vanderBij:1981,Unruh:1988,Ng:1990,Ellis:2010} . In our paper, we focus on the action that gives rise to volume preserving diffeomorphsims \cite{Buchmuller:1988,Padilla:2014}. 
\begin{equation} \label{eq:action-functional-UG}
   S_{UG}= \frac{2}{\kappa^2}\bigg[ \int \sqrt{-g} \, d^4x\,\, \{R-2\lambda(x)\} + \int f\, d^4x\,\, 2\lambda(x)\bigg] + S_M [g_{\mu\nu}, \Psi].
\end{equation}
    In Eq.(\ref{eq:action-functional-UG}), $\kappa\equiv \sqrt{32\pi G}$, $R$ is the Ricci scalar, $\lambda(x)$ denotes the Lagrange multiplier. In the second integral, $f$ is the non-dynamical volume element. The matter lagrangian is given by $\mathcal{L}_M$, having $\Psi$ as matter fields. The field equations resulting from the variation of $g_{\mu\nu}$, $\lambda$ and $\Psi_M$ yields
    \begin{gather}
       R_{\mu\nu}-\frac{1}{2}R\,g_{\mu\nu}+\lambda(x) g_{\mu\nu}= \frac{\kappa^2}{4} T_{\mu\nu}, \label{eq:metric-variation}\\
       f=\sqrt{-g}, \label{eq:lagrange-multiplier-variation}\\
       \df{\delta S_M}{\delta \Psi}=0,
    \end{gather}
where, 
\begin{equation}
    T_{\mu\nu}=-\dfrac{2}{\sqrt{-g}}\dfrac{\delta S_M}{\delta g^{\mu\nu}}. 
\end{equation}
Eq.(\ref{eq:lagrange-multiplier-variation}) shows that the volume form is fixed in UG.  This is achieved by varying the Lagrange multiplier $\lambda(x)$. We will see later how this fixed volume form results in a restricted class of diffeomorphisms in UG. To remove $\lambda$,  we trace Eq.(\ref{eq:metric-variation}) and find $\lambda=\frac{1}{4}(\frac{\kappa^2}{4} T+R)$. Thus, the field equation in UG take the tracefree part of the Einstein field equation,
\begin{equation} \label{eq:UG-tracefree}
    R_{\mu\nu}-\frac{1}{4}R\, g_{\mu\nu}= \frac{\kappa^2}{4} \bigg(T_{\mu\nu}-\frac{1}{4}g_{\mu\nu}T\bigg)
\end{equation}

\subsection{Violation of EM tensor conservation in UG} \label{sec:EM-violation}

\noindent Now we describe how non-conservation of energy momentum tensor arises in UG. We draw upon a comparative analysis by first illustrating the case in GR.

In GR, matter action $S_M$ must be invariant under any general coordinate transformation or diffeomorphism invariance. Let $\xi$ be the vector field associated to diffeomorphisms, then
\begin{equation} \label{eq:matter_action_variation}
    0=\delta_\xi \, S_M= \int \df{\delta S_M}{\delta g^{\mu\nu}} \,\, \delta_\xi  g^{\mu\nu} + \int \df{\delta S_M}{\delta \Psi}\,\, \delta_\xi\Psi.
\end{equation}
Let us consider that the matter field equations are satisfied, $\df{\partial S_M}{\partial \Psi} \bigg|_\Psi=0$. Thus, Eq.\eqref{eq:matter_action_variation} becomes 
\begin{equation} \label{eq:derivation_EOM_GR}
    0= \int \df{\delta S_M}{\delta g^{\mu\nu}} \,\, \delta_\xi  g^{\mu\nu} = \int \sqrt{-g} \, T_{\mu\nu}  \, \nabla^{(\mu} \xi^{\nu)} \, d^4x =-\int \sqrt{-g} \, \,\, (\nabla^{\mu} T_{\mu\nu}) \,\xi^{\nu}  \, d^4x
\end{equation}
where in Eq.(\ref{eq:derivation_EOM_GR})  we have used the following definitions.
\begin{equation} \label{eq:definition}
  \delta_\xi g_{\mu\nu} = 2 \nabla_{(\mu} \xi_{\nu)}   
\end{equation}
The definition of the EM tensor from the variation of the matter action have been mentioned earlier. The last equality in Eq.(\ref{eq:derivation_EOM_GR}) can be arrived by setting the surface integral to zero. Thus, for general diffeomorphism invariant action, $T_{\mu\nu}$ is always conserved by virtue of the matter field equations. 
\begin{equation} \label{eq:EM_tensor_cons}
    \nabla^\mu T_{\mu\nu} = 0. 
\end{equation}

Next, let us treat case of UG. Recall Eq.(\ref{eq:lagrange-multiplier-variation}), which resulted in a fixed volume form. Thus, in this scenario, we consider volume preserving diffeomorphisms, {\em i.e.} $\delta_\xi g=0$ where $g$ is the determinant of the metric tensor. This condition can be written as
\begin{equation} \label{eq:volume_preserving_diff}
    \delta_\xi \, g= g\, g^{\mu\nu} \, \delta_\xi g_{\mu\nu} = g^{\mu\nu} \delta_\xi g_{\mu\nu} =0.
\end{equation}
Utilising Eq.(\ref{eq:definition}), we get 
\begin{equation} \label{eq:transverse_diff}
    g^{\mu\nu} (\nabla_\mu \, \xi_\nu + \nabla_\nu \, \xi_\mu) = \nabla^\mu\, \xi_\mu =0 
\end{equation}
Hence, the class of volume preserving differmorphism for non-degenerate metric fields turn out to be transverse diffeomorphisms. Since $\nabla_\mu\, \xi^\mu =0$, $\xi^\mu$ can be written in terms of an antisymmetric rank-2 tensor field ($\omega_{\alpha\beta}$).
\begin{equation} \label{eq:xi_definition}
    \xi_\mu=\epsilon_{\mu\nu\alpha\beta}\, \nabla^{\nu}\,  \omega^{\alpha\beta}
\end{equation}
Then, from Eq.(\ref{eq:derivation_EOM_GR}) we find that,
\begin{align} \label{eq:UG_EM}
       0= & \int \sqrt{-g} \, \, (\nabla^{\mu} T_{\mu\nu}) \,\epsilon^{\nu\sigma\alpha\beta}\, \nabla_{\sigma}\,  \omega_{\alpha\beta}\, d^4x = \int \sqrt{-g} \, \, \nabla_{\sigma}\,(\nabla^{\mu} T_{\mu\nu} \,\epsilon^{\nu\sigma\alpha\beta})\,   \omega_{\alpha\beta} \, d^4x\nonumber \\
      & = \int \sqrt{-g} \, \, (\nabla_{\sigma}\,J_{\nu} )\,\epsilon^{\nu\sigma\alpha\beta}\,   \omega_{\alpha\beta}\, d^4x
\end{align}
Thus, Eq.(\ref{eq:UG_EM}) implies $\bold{d} J=0$, {\em i.e.} the exterior derivative of the one-form $J$ vanishes where $J_\nu:= \nabla^\mu T_{\mu\nu}$. Since $J$ is a closed form, locally, it can be expressed as an exact form $J= \bold{d} Q$. 
\begin{equation} \label{eq:nonconservation_UG}
    \bold{d}^2 \, Q =0 \Rightarrow \nabla^\mu\, (T_{\mu\nu}-g_{\mu\nu} Q) =0.
\end{equation}
 $Q$ is a scalar function signifying the violation of the EM tensor and is termed as the {\em diffusion} parameter. Introducing $\nabla^a$ in Eq.(\ref{eq:metric-variation}) and using Eq.(\ref{eq:nonconservation_UG}) gives, 
 \begin{equation}
     \lambda= \Lambda_0+ \frac{\kappa^2}{4} Q(x) 
     \label{eq: diffusion func}
 \end{equation}
$\Lambda_0$ appears as a integration constant which can be treated as a cosmological constant. There have been efforts to define an effective cosmological constant ($\Lambda_0+ \frac{\kappa^2}{4} Q$) by fine-tuning the integration constant to solve the cosmological constant problem \cite{Perez:2017-PRL,Bengochea:2019,PhysRevLett.118.021102}. Using the above construction of naturally generating the cosmological constant along with diffusion, there have been efforts to alleviate the Hubble tension \cite{Perez:2020-H0}, and also give strong cosmological constraints in UG \cite{Landau:2022}. In an apparently different context, the low spin found in black holes from GW observations has been explained through this UG paradigm involving diffusion \cite{Perez:2019-BH}.

Note that if $Q(x)$ is constant, one gets back the same Einstein field equation of GR. For the purpose of calculations, we will not be considering the impact of cosmological constant since we are studying the GW radiation coming from astrophysical binary systems. 

% We note that the Newtonian potential in UG is is identical to that in Einstein's gravity which can be see as follows.  The potential between non-relativistic bodies on exchange of a graviton is given in UG by 
% \be
% V(k)=\frac{\kappa^2}{2} T_1^{00}\,\frac{1}{4} \frac{1}{k^2} \,T_2^{00} = 4 \pi^2 G M^2 \frac{1}{k^2}
% \ee
%  which in position space is $V(r)= G M_1 M_2/r$ which is the Newtonian potential. UG will gives different prediction compared to Einstein's theory of GR when all components of the stress tensor are probed as in the case of gravitational waves where the relation between the $T_{ij}$ and $T_{00}$ components is different compared to Einsteins gravity because of the difference in the energy momemtum conservation equation  (\ref{eq:nonconservation_UG}). 

EM tensor conservation violation is also used in Rastall gravity \cite{Rastall:1972}. In fact, it can be shown that Eq.(\ref{eq:metric-variation}) is identical to field equations of Rastall gravity in a certain case \cite{Visser:2017}. However, the EM tensor violation in Rastall gravity is ad-hoc and  unlike, UG, does not arise from a deeper geometrical principle. In UG, the trace-free condition on the energy-momentum tensor follows naturally from the fixed-volume constraint on the metric determinant, providing a more theoretically motivated foundation.

\section{Gravitational radiation}
\label{sec:radiation}
In the previous section, we have shown how the presence of a reduced set of diffeomorphisms, called transverse diffeomorphisms, gives rise to a violation in the  conservation of EM tensor. In this section we will see how this impacts the energy radiated between a binary system  following the prescription given in \cite{Mohanty:2020pfa,Poddar2022}.

 In GR, to obtain the interaction vertex, one starts by varying the matter action in the following way, taking $g_{\mu\nu}=\bar{g}_{\mu\nu}+\kappa h_{\mu\nu}$.
\begin{align}
    \delta S_M= & \int \, d^4x\,  \delta (\sqrt{-g} \mathcal{L}_M) \nonumber \\
    = & \int \, d^4x\, \sqrt{-g} \bigg( \frac{\delta \mathcal{L}_M}{\delta g_{\mu\nu}}- \frac{1}{2} \mathcal{L}_M g^{\mu\nu}\bigg) \bigg|_{g_{\mu\nu}=\bar{g}_{\mu\nu}} \,\, \delta g_{\mu\nu} \nonumber \\
    & = \int \, d^4x\, \sqrt{-\bar{g}} \, \, \bigg(-\frac{\kappa}{2} T^{\mu\nu}\, h_{\mu\nu}\bigg).
\end{align}
Hence, the interaction vertex is $\frac{\kappa}{2}\, h_{\mu\nu}\, T^{\mu\nu}$ in GR. In the UG case, in Appendix \ref{app:matter-graviton} we have shown how transverse diffeomorphisms keeps the interaction term invariant.

Therefore, the emission rate of massless gravitons (gravitational radiation) from this vertex becomes \cite{Mohanty:2020pfa,Poddar2022},
\begin{eqnarray}
 \label{eq:emission-rate}
    d\Gamma &=& \frac{\kappa^2}{4}  \sum_{\lambda=1}^{2} \lvert T_{\mu\nu} (k^\prime) \, \epsilon_\lambda^{\mu\nu}(k^\prime) \rvert^2 \, (2\pi) \delta (\omega-\omega^\prime) \frac{d^3k}{(2\pi)^3}  \, \frac{1}{2\omega} \nonumber \\
 &=& \frac{\kappa^2}{8(2\pi)^2}  \sum_{\lambda=1}^{2} [ T_{\mu\nu} (k^\prime) \,  T_{\alpha\beta}^* (k^\prime) \epsilon_\lambda^{\mu\nu}(k^\prime) \epsilon_\lambda^{*\,\mu\nu}(k^\prime) \,]  \delta (\omega-\omega^\prime) \frac{d^3k}{2\omega} 
\end{eqnarray}

In Eq.(\ref{eq:emission-rate}), the polarization tensor sum for spin-2 massless gravitons yields the same as that for GR \cite{Henneaux:1989}
\begin{equation} \label{eq:polarization-sum}
    \sum_{\lambda=1}^{2}\, \epsilon^\lambda_{\mu\nu} \,\, \epsilon^{*\lambda}_{\alpha\beta}= \frac{1}{2} (\eta_{\mu\alpha}\, \eta_{\nu\beta} + \eta_{\mu\beta}\, \eta_{\nu\alpha} - \eta_{\mu\nu} \eta_{\alpha\beta}).
\end{equation}
The polarization sum equivalent to GR can also be understood from another argument. Eq.(\ref{eq:polarization-sum}) follows from the massless Fierz-Pauli action. This action can be derived by taking the expanding Einstein-Hilbert action around Minkowski metric till second order. The derivation does not require the conservation of EM tensor. Hence, as UG only differs from GR through violation of the conserved EM tensor, the Fierz-Pauli action is restored in that scenario.

The energy radiation rate ($dE/dt$) due to the emission of a massless graviton from this vertex is related to the emission rate as $dE/dt = \int \omega\, d\Gamma$. Thus, the energy radiation rate can be expressed in terms of the energy-momentum tensor of the source as,
\begin{equation} \label{eq:energy-loss}
\frac{dE}{dt}=\frac{\kappa^2}{8\, (2\pi)^2} \int \bigg[\lvert T_{\mu\nu} (k^\prime)\rvert^2-\frac{1}{2}\lvert T^{\mu}\,_\mu (k^\prime)\rvert^2\bigg]\, \delta (\omega-\omega^\prime)\, \omega^2 \, d\omega \, d\Omega_k ,
\end{equation}
where `$\prime$' denotes that the rate of energy loss is evaluated at the mode $k^\prime_{\mu}=(\vec{k}^\prime,\omega^\prime)$ and differentiates it from the integration variable $k_{\mu}=(\vec{k},\omega)$.

Due to the non-conservation of $T_{\mu\nu}$, we find the dispersion relation is also modified. Instead of $k^\mu \, T_{\mu\nu}=0$ we have,
\begin{equation} \label{eq:dispersion-relation}
    k^\mu (T_{\mu\nu}-g_{\mu\nu}Q)=0.
\end{equation}
Taking $k^\mu=(\omega,\vec{k})$, $\lvert\vec{k}\rvert=\omega$ and defining $\hat{k}^\mu=\frac{k^\mu}{\omega}$ we find,
\begin{align}
    T_{0i}=-\hat{k}^j\, T_{ji} + \hat{k}_i\, Q, \label{eq:T_0i} \\
    T_{00}=-2Q + \hat{k}^i\, \hat{k}^j\, T_{ij}. \label{eq:T_00}
\end{align}
Thus, the integrand inside square braces in Eq.(\ref{eq:energy-loss}) gets modified as
\begin{align} \label{eq:Stress-Energy-tensor-modification}
    \bigg[\lvert T_{\mu\nu} (k^\prime)\rvert^2-\frac{1}{2}\lvert T^{\mu}\,_\mu (k^\prime)\rvert^2\bigg]  = &  \Lambda_{ij,lm}\, T^{ij*} \, T^{lm} 
    -Q^* \, \delta_{ij}\, T^{ij} - Q \, \delta_{ij}\, T^{ij*} \nonumber \\
      & + Q^*\, \hat{k}^i\, \hat{k}^j\, T^{ij} + Q\, \hat{k}^i\, \hat{k}^j\, T^{ij*}  
\end{align}
 where, $\Lambda_{ij,lm}= \delta_{il}\, \delta_{jm} - 2 \hat{k}_j \hat{k}_m \delta_{il}+\frac{1}{2} \hat{k}_i\hat{k}_j\hat{k}_l\hat{k}_m -\frac{1}{2} \delta_{ij} \delta_{lm}+\frac{1}{2} (\delta_{ij}\hat{k}_l\hat{k}_m +\delta_{lm}\hat{k}_i\hat{k}_j)$.
Note in Eq.(\ref{eq:Stress-Energy-tensor-modification}) the limit $Q=0$ gives back the GR formula \cite{Poddar2022}. The angular integral  becomes,
\begin{align} \label{eq:Stress-Energy-tensor-modification-angular-integral}
     \int  \bigg[\lvert T_{\mu\nu} (k^\prime)\rvert^2-\frac{1}{2}\lvert T^{\mu}\,_\mu (k^\prime)\rvert^2\bigg] \, d\Omega_k = & \frac{8\pi}{5} \bigg(\lvert T_{ij} (k^\prime)\rvert^2-\frac{1}{3}\lvert T^{i}\,_i (k^\prime)\rvert^2 \bigg) \nonumber\\
   & -\frac{8\pi}{3} \bigg(  Q^*(k^\prime)\, \delta_{ij} \, T^{ij}(k^\prime) + Q(k^\prime)\, \delta^{ij} \, T_{ij}^*(k^\prime) \bigg)
\end{align}
For details of angular integrals and their formulas, the interested reader can look up \cite{Poddar2022}. 
The stress energy  tensor for the binary system is given as,
\begin{equation}
    T_{\mu\nu} = \mu\, \delta^3(\vec{x^\prime}-\vec{x}(t))\, U_\mu\, U_\nu \label{eq:binary-pulsar-EM}.
\end{equation}
Here, $\mu=\frac{M_1 M_2}{M_1+M_2}$ is the reduced mass of the binary system and $M_1, M_2$ represent individual masses of the binary, $\vec{x}(t)$ is the binary orbit and the four-velocity of the binary is $U_\mu =(1, \dot{x}, \dot{y}, 0)$. In parametric form, the Keplerian orbits can be expressed as,
\begin{equation} \label{eq:orbit-parametric-form}
    x= a (\cos \xi-e), \hspace{1cm} y= a \sqrt{1-e^2} \sin\xi, \hspace{1cm} \Omega\, t=\xi-e\sin\xi 
\end{equation}
where $a, e$ are the semi-major axis and the eccentricity of the orbit, respectively. The Fourier transformed velocity components are,
\begin{equation} \label{eq:fourier-transform-velocity}
    \dot{x}_n =\frac{1}{T}\int_0^T\, e^{in\Omega t} \dot{x} \, dt= - i\, a\, \Omega\, J_n^\prime (ne), \hspace{0.5cm} \dot{y}_n= \frac{1}{T}\int_0^T\, e^{in\Omega t} \dot{y} \, dt = \frac{a\sqrt{1-e^2}}{e}\,  \Omega\, J_n (ne)
\end{equation}
where $T=\frac{2\pi}{\Omega}$ is the orbital period and $J_n(z)=\frac{1}{2\pi}\int_0^{2\pi}\, d\xi\,e^{i(n\xi-z\,\sin\xi)}$ is the Bessel function identity. From the modified conservation equations (\ref{eq:T_0i}) and (\ref{eq:T_00}) we get,
\begin{equation} \label{eq:modified-dispersion-relation}
    \partial^i\, \partial^j\, T_{ij} ({\bf x}, \omega^\prime) = - {\omega^\prime}^2 [T_{00}(\vec{x}, \omega^\prime)\,+2Q (\vec{x}, \omega^\prime)] 
\end{equation}
where $\omega^\prime=n\Omega$. Eq.(\ref{eq:modified-dispersion-relation}) can be re-expressed in the following,
\begin{equation} \label{eq:modified-dispersion-relation-2}
    T_{kl}(\omega^\prime)=-\frac{{\omega^\prime}^2}{2}\int \, d^3x\, [T_{00}(\vec{x}, \omega^\prime)\,+2Q (\vec{x}, \omega^\prime)] \, x^\prime_k\,x^\prime_l.
\end{equation}
In order to proceed further, we consider $Q$ to be of the form, 
\begin{equation} \label{eq:diffusion-parameter-form}
    Q(x^\prime)= \zeta \, \mu \, \delta^3(\vec{ x^\prime}-\vec{x}(t)).
\end{equation}
where $\zeta$ is the dimensionless coupling constant signifying the coupling between the binary system with the diffusion parameter. Utilising the functional form of Eq.(\ref{eq:diffusion-parameter-form}) and using the expressions for the orbit from Eq.(\ref{eq:orbit-parametric-form}) and substituting it in Eq.(\ref{eq:modified-dispersion-relation-2}) we get 
\begin{gather} 
     T_{xx} (\omega^\prime)= -\frac{(1+2\zeta) \mu {\omega^\prime}^2\, a^2}{4n} [J_{n-2} (ne) -2 e J_{n-1} (ne)+2 e J_{n+1} (ne) -J_{n+2}(ne)] \label{eq:T_xx-expressions},\\
    T_{yy} (\omega^\prime) = \frac{(1+2\zeta) \mu {\omega^\prime}^2\, a^2}{4n} [J_{n-2} (ne) -2 e J_{n-1} (ne)+2 e J_{n+1} (ne) -J_{n+2}(ne)+\frac{4}{n} J_n (ne)] \label{eq:T_yy-expressions},\\
    T_{xy} (\omega^\prime) = -i\frac{(1+2\zeta) \mu {\omega^\prime}^2\, a^2}{4n} \sqrt{1-e^2} [J_{n-2} (ne)+ J_{n+2}(ne) -2 J_n(ne)] \label{eq:T_xy-expressions}.
\end{gather}
From Eqs.(\ref{eq:T_xx-expressions}), (\ref{eq:T_yy-expressions}) and (\ref{eq:T_xy-expressions}) we find that,
\begin{gather}
    T_{ij} (\omega^\prime) \, T^{ij*} (\omega^\prime)= 4 (1+2\zeta)^2 \mu^2 {\omega^\prime}^4 a^4\bigg[f(n,e) + \frac{J_n^2(ne)}{12n^4}\bigg], \label{eq:T_ij-squared}\\
    \lvert T^i\,_i\rvert^2 = \frac{\mu^2 {\omega^\prime}^4 a^4}{n^4} \, J_n^2(ne). \label{eq:T_ij-squared-mod}
\end{gather} 
 where, 
\begin{eqnarray}
f(n,e)=\frac{1}{32n^2}\bigg\{ \bigg[J_{n-2} (ne) -2 e J_{n-1} (ne)+2 e J_{n+1} (ne) -J_{n+2}(ne)+\frac{2}{n} J_n (ne)\bigg]^2    \nonumber \\
    +(1-e^2) [J_{n-2} (ne)+ J_{n+2}(ne) -2 J_n(ne)]^2 +\frac{4}{3n^2} J_n^2 (ne)\bigg\}
    \end{eqnarray}
To evaluate the integral given in Eq.(\ref{eq:energy-loss}), we need to evaluate $Q(k^\prime)$ (see Eq.(\ref{eq:Stress-Energy-tensor-modification-angular-integral})). To go about this, we use similar technique as was used to calculate the fourier transformed velocity components of the binary orbit in Eq.(\ref{eq:fourier-transform-velocity}). 
\begin{eqnarray}\label{eq:Q-fourier-domain}
     Q(\vec{k^\prime}, \omega^\prime) = & \dfrac{\zeta \mu}{T} \displaystyle\int_0^T \displaystyle\int \delta^{(3)} (\vec{x^\prime}-\vec{x}(t))\, e^{-i (\vec{k^\prime}\cdot \vec{x}-\omega^\prime t)} \, d^3x \, dt \nonumber \\
     & =\dfrac{\zeta \mu}{T} \displaystyle\int_0^T  e^{-i (\vec{k^\prime}\cdot \vec{x^\prime}(t)-\omega^\prime t)} \, dt =Q(\omega^\prime)
\end{eqnarray}
Assuming that the length scale of the fourier modes is large compared to the dimensions of the orbit, $\frac{1}{\lvert\vec{k^\prime}\rvert}>>a$, 
we can Taylor expand the terms in Eq.(\ref{eq:Q-fourier-domain}). Keeping till the quadratic order,  we find
\begin{equation} \label{eq:Q-fourier-domain-2}
    Q(\omega^\prime) =\frac{\zeta\mu}{T} \displaystyle\int_0^T \bigg[1-i\,\vec{k^\prime}\cdot\vec{x^\prime}(t) - \frac{1}{2} (\vec{k^\prime}\cdot\vec{x^\prime}(t))^2\bigg]\, e^{i\omega^\prime t }\, dt 
\end{equation}
After performing the integration, Eq.(\ref{eq:Q-fourier-domain-2}) can be written as,
\begin{equation}
    Q(\omega^\prime) = \zeta \mu \bigg\langle -i (k_x^\prime \, x^\prime(\omega^\prime)+k_y^\prime \, y^\prime(\omega^\prime))-\frac{1}{2} \bigg[(k_x^\prime \, x^\prime(\omega^\prime))^2+(k_y^\prime \, y^\prime(\omega^\prime))^2 + 2k_x^\prime k_y^\prime x^\prime(\omega^\prime) y^\prime(\omega^\prime)\bigg] \bigg\rangle
\end{equation}
 We find that the mean wavevector integrals over all solid angles give, $\langle k_x^\prime\rangle=\langle k_y^\prime\rangle=\langle k_x^\prime \, k_y^\prime\rangle=0$ and $\langle {k_x^\prime}^2\rangle=\langle {k_y^\prime}^2\rangle=\frac{\Omega^2}{3}$, where $k_x^\prime= \vert \vec{k^\prime} \vert \sin \theta \cos \phi$, $k_y^\prime= \vert \vec{k^\prime} \vert \sin \theta \sin \phi$, $k_z^\prime= \vert \vec{k^\prime} \vert \cos \theta$ and the mean value $<f(\vert\vec{k}^\prime\vert,\theta,\phi)>=\frac{1}{4\pi} \int f(\vert\vec{k}^\prime\vert,\theta,\phi) d\Omega = \frac{1}{4\pi} \int_{\theta=0}^{\pi}\int_{\phi=0}^{2\pi} f(\vert\vec{k}^\prime\vert,\theta,\phi) \sin \theta d\theta d\phi$ . Therefore, effectively,    
\begin{eqnarray} \label{eq:Q-final-form}
Q(\omega^\prime)= - \frac{\zeta\mu a^2}{3} (n\Omega)^2\, J_n(ne).
\end{eqnarray}

Using Eqs.(\ref{eq:T_ij-squared}), (\ref{eq:T_ij-squared-mod}) and (\ref{eq:Q-final-form}) we get the final expression for the energy loss as,

\begin{equation}
    \frac{dE}{dt} = \frac{32 G}{5} \mu^2 a^4 \Omega^6 \bigg[ (1+2\zeta)^2\, f(e) + \frac{5\zeta}{18} (1+2\zeta)\, g(e) \bigg]
    \label{dEbydt_UG}
\end{equation}

where,
\begin{align}
    f(e)= \dfrac{1+\frac{73}{24}e^2+\frac{37}{96}e^4}{(1-e^2)^{7/2}} \hspace{2cm} g(e)=\dfrac{e^2+\frac{e^4}{4}}{4(1-e^2)^{7/2}}.
\end{align}

\section{Observational constraints}
\label{sec:obs}

In this section, using the formulae derived in the previous section, we compare the observed
period decay of the compact binary systems (neutron star-neutron star/white dwarf) with
that predicted theoretically in UG. We constrain the dimensionless parameter $\zeta$ in the theory  and compare the constraint with the astrophysical constraint \cite{Yang:2022ees}. Six binary objects are used for our analysis, i.e., PSR B1913+16 (Hulse-Taylor binary) \cite{Weisberg:2016jye}, PSR
J1141-6545 (NS-WD binary) \cite{Bhat:2008ck}, PSR J0735-3039 (double pulsar) \cite{Kramer:2006nb}, PSR B2127+11C \cite{2006ApJ...644L.113J}, PSR B1534+12 \cite{Stairs_2002}, and PSR J1756-2251 \cite{Ferdman:2014rna} listed in table~\ref{tab:objects}.
We define relative ‘change’ of intrinsic (i.e. observed) orbital period decay and UG predicted orbital period decay with respect to that predicted from GR as,
\begin{eqnarray}
\Delta_{\text{Obs}}&=&\left|\frac{\dot{P}_{b,\text{Intrinsic}}-\dot{P}_{b,\text{GR}}}{\dot{P}_{b,\text{GR}}}\right|  \label{Delta-obs},\\
\Delta_{\text{UG}}&=&\left|\frac{\dot{P}_{b,\text{UG}}-\dot{P}_{b,\text{GR}}}{\dot{P}_{b,\text{GR}}}\right|.
\label{Delta-UG}
\end{eqnarray}
The UG model must satisfy the condition $\frac{\Delta_{UG}}{\Delta_{Obs}}<1$. 

\begin{table}[h]
\centering
\resizebox{\linewidth}{!}{%
\begin{tabular}{ |l|c|c|c|c|c|c|c| }
 
 \hline
Parameters \hspace{0.01cm} & \textbf{PSR B1913+16}\hspace{0.01cm}&\textbf{PSR J1141-6545}\hspace{0.01cm}&\textbf{PSR J0735-3039}\hspace{0.01cm}&\textbf{PSR B2127+11C}\hspace{0.01cm}&\textbf{PSR B1534+12}\hspace{0.01cm}&\textbf{PSR J1756-2251}\hspace{0.01cm}\\
 \hline
\textbf{Pulsar mass} $m_1$ ($M_{\odot}$) &$1.438\pm 0.001$ &$1.27\pm0.01$ & $1.3381$ & $1.358(10)$ & $1.333228 $ & $1.341 $\\
\textbf{Companion mass} $m_2$ ($M_{\odot}$)&$1.390\pm 0.001$ & $1.02\pm0.01$ & $1.2489$ & $1.354(10)$ & $1.3452(10) $ & $1.230 $\\
\textbf{Eccentricity} $e$ &$0.6171340(4)$ & $0.171884(2)$ & $0.0877775$ & $0.681395(2)$ & $ 0.2736775(3)$ & $0.1805694$\\
\textbf{Orbital period} $P_b$ (d)&$0.322997448918(3)$& $0.1976509593(1)$ & $0.10225156248$ & $0.33528204828(5)$ & $ 0.420737299122$ & $0.31963390143$\\
\textbf{Intrinsic} $\dot{P_b}(10^{-12}\rm{ ss^{-1}})$ &$-2.398\pm 0.004$ &$-0.403(25)$ &$-1.252(17)$ & $ -3.95(13)$ & $-0.137(3) $ & $-0.229 $ \\

 \hline
\end{tabular}
}
\caption{Summary of the measured orbital parameters and the orbital period derivative values from observation and GR for PSR B1913+16 (Hulse-Taylor binary) \cite{1975-Hulse-ApJ,Weisberg:2016jye}, PSR
J1141-6545 (NS-WD binary) \cite{Bhat:2008ck}, PSR J0735-3039 (double pulsar) \cite{Kramer:2006nb}, PSR B2127+11C \cite{2006ApJ...644L.113J}, PSR B1534+12 \cite{Stairs_2002}, and PSR J1756-2251 \cite{Ferdman:2014rna}. The uncertainties in the last digits are quoted in the parenthesis.}
\label{tab:objects}
\end{table}

We note that the Newtonian potential in UG is identical to that in Einstein's gravity which can be seen as follows.  The potential between non-relativistic bodies on exchange of a graviton is given in UG by 
\be
V(\vec{k})=\frac{\kappa^2}{2} T_1^{00}\,\frac{1}{4} \frac{1}{\vec{k}^2} \,T_2^{00} = 4 \pi^2 G M^2 \frac{1}{\vec{k}^2}
\ee
 which in position space is $V(r)= G M_1 M_2/r$ which is the Newtonian potential and $\vec{k}$ is the spatial components of the four momentum $k^\mu$. UG will give different prediction compared to Einstein's theory of GR when all components of the stress tensor are probed as in the case of gravitational waves where the relation between the $T_{ij}$ and $T_{00}$ components is different compared to GR because of the difference in the EM conservation equation  (\ref{eq:nonconservation_UG}). 

Thus  Newtonian gravitational force  between the binary stars is unchanged and thus the orbital period decay in UG is given by,
\begin{equation}
    \dot{P}_{b}=- 6 \pi G^{-3/2}(M_{1}M_{2})^{-1}(M_{1}+M_{2})^{-1/2}a^{5/2}\dot{E}
    \label{pbdot }
\end{equation}
where $\dot{E}$ is given by Eq.~(\ref{dEbydt_UG}).  The orbital period decay in GR is given by the Peters-Mathews formula \cite{Peters:1963ux}. Comparing theoretically predicted value with the observation we obtain constraint on the dimensionless parameter $\zeta$  which are given in Table~\ref{tab:constraint} 

%\begin{table}[t]
%\centering
%\begin{tabular}{ |l|c| }
 
% \hline
%Binary system \hspace{0.01cm} & $\vert \zeta\vert$\hspace{0.01cm}\\
% \hline
%PSR B1913+16 & $\leq 5\times 10^{-4}$ \\
%PSR J1141-6545 & $\leq 0.01$ \\
%PSR J0735-3039 & $\leq 3\times 10^{-3}$\\
% PSR B2127+11C & $\leq 6\times 10^{-4}$ \\
% PSR B1534+15 & $\leq 0.067$ \\
% PSR J1756-2251 & $\leq 0.014$\\
%  \hline
% \end{tabular}
% \caption{Observational constraint on $\zeta$ from binary systems.}
% \label{tab:constraint}
% \end{table}

\begin{table}[h]
\centering
\begin{tabular}{l@{\hspace{2cm}}l}
\hline
Binary system & $\vert \zeta\vert$ \\
\hline
PSR B1913+16 & $\leq {\bf 5\times 10^{-4}}$ \\
PSR J1141-6545 & $\leq 0.01$ \\
PSR J0735-3039 & $\leq 7\times 10^{-4}$\\
PSR B2127+11C & $\leq 6\times 10^{-4}$ \\
PSR B1534+15 & $\leq 0.067$ \\
PSR J1756-2251 & $\leq 0.014$\\
\hline
\end{tabular}
\caption{\label{tab:constraint} Observational constraint on $\zeta$ from binary systems.}
\end{table}

The most stringent constraint comes from the binary pulsar systems: PSR B1913+16 (Hulse-Taylor) and PSR B2127+11C. The constraint on $\zeta$ comes out to be $\vert \zeta \vert \leq 5\times 10^{-4}$. Also note that the dimensionless parameter $\zeta$ can have positive or negative signature. Let us compare our obtained constraint on the non-conservation of the EM tensor in UG with that obtained in \cite{Yang:2022ees}, where the authors studied the tidal deformability of neutron star in the context of the gravitational wave observations, particularly for GW170817 and GW190425. In their work, the authors assumed that the non-conservation behavior of EM tensor takes the form,
$\nabla^{\mu}T_{\mu\nu}=\alpha \rho \delta^r_{\nu}$, where $\alpha$ is the non-conservation parameter which have the dimension of the inverse of the length and $\rho$ is the mass density of the neutron star. This is different from our case where we have $\nabla^{\mu}T_{\mu\nu}=\zeta \nabla_{\nu} \rho_m$, as we assumed $Q(x)=\zeta \rho_m$ (see Eq.~(\ref{eq:diffusion-parameter-form})) and $\rho_m= \mu \, \delta^3(\vec{ x^\prime}-\vec{x}(t))$ is the matter density of the binary star system. However, assuming a typical size of a neutron star $R\sim 10 $ km, we can  qualitatively compare the two non-conservation parameters $\zeta$ and $\alpha$ by a relation $\zeta\equiv  \alpha R$.  From the tidal deformability of the neutron stars, the authors obtained a constraint on $\alpha$ as $\vert \alpha \vert  \leq 0.02/\left(1.73\times 10^6 cm\right)= 1.15607\times 10^{-8}\text{cm}^{-1}$  \cite{Yang:2022ees}. This translates an equivalent constraint on $\zeta$ as $\vert \zeta\vert \leq 0.0116$, where as we obtain the constraint from the orbital period decay of the binary pulsar systems as $\vert \zeta \vert \leq 5\times 10^{-4}$ . Thus our constraint on the non-conservation parameter $\zeta$ is better by at least one order of magnitude.

Earlier in Section II, we have already mentioned (Eq.~\ref{eq: diffusion func}) that an effective cosmological constant, which actually varies with time, emerges out of the diffusion function $Q(x)$. In the cosmological scenario, the non-conservation of the EM tensor leading to the emergence of such effective cosmological constant can arise due to different reasons, such as nonunitary modifications of quantum dynamics, in causal set approach to quantum gravity, etc. \cite{PhysRevLett.118.021102}. For example, the effective cosmological constant induced by a wave function collapse of baryons, using the mass-proportional continuous spontaneous localization (CSL) model was discussed in \cite{PhysRevLett.118.021102}. The CSL model provides a microscopic origin of the non-conservation of the energy-momentum tensor, which is completely different from the macroscopic model as in our case and thus these are not comparable. There are other phenomenological studies in cosmology, such as resolving the Hubble tension \cite{Perez:2020-H0}, study of the modifications of the predictions for the anisotropy and polarization of the Cosmic
Microwave Background (CMB) \cite{Landau:2022}, etc., which differentiate GR from UG, but definitive constraints on the non-conservation parameter are absent.

\section{Conclusions}\label{sec:conclusions}

In this article, we have studied the phenomenology of UG in the context of gravitational radiation emitted from binary systems. Signatures of UG only differ from GR, classically, when the conservation of EM tensor is violated. Considering such non-conservation introduces a dimensionless parameter $\zeta$, responsible for the violation of conserved EM tensor.  The main aim of this work has been to constrain this theory parameter $\zeta$ from astrophysical binary pulsar data.  

%To begin with, we have demonstrated how the unimodular constraint of volume preserving diffeomorphisms in UG gives rise to transverse diffeomorphisms.  Subsequently, we considered an action which keeps the volume form fixed by introducing a Lagrange multiplier. Considering such an action, we have shown how the field equations are modified in UG {\em vis-a-vis} GR. We find that there is a violation of EM tensor due to nonzero diffusion function $Q$. A interesting feature of the resulting field equation is that an integration constant results due to the non-conservation 

To achieve this, we have started with an action involving a Lagrange multiplier whose variation gives rise to a fixed volume form. Variation w.r.t. the metric field and then removing the Lagrange multiplier yields the tracefree Einstein field equations. Subsequent to this, we have analysed the status of conservation of EM tensor in UG having volume preserving diffeomorphisms {\em vis-a-vis} GR.  In case of GR, the full diffeomorphism invariant action gives rise to conservation of EM tensor due to the matter field equations.  
For UG, we find that for non-degenerate metric fields, volume preserving diffeomorphisms turn out to be the transverse diffeomorphisms. This restricted class of diffeomorphisms acting on the matter action shows that divergence of EM tensor is proportional to the derivative of a scalar field, coined as the diffusion function. The Lagrange multiplier turns out to contain the diffusion function and an integration constant, which can be taken as the cosmological constant. As our work pertained to astrophysical binaries, we set the integration constant to zero.

Having set the stage by choosing to work with UG, we next set out to compute the gravitational radiation emitted by binary  systems in this theory. As UG has only two propagating DOF, the polarization sum remains the same as that of GR. Gravitational radiation in this scenario consists of emission of massless gravitons from the non-conserved matter given by $T_{\mu\nu}$. However, in Appendix \ref{app:matter-graviton}, we have shown how the interaction term remains invariant even when EM tensor conservation is violated for transverse diffeomorphisms. However, due to the violation, there are quantitative differences in the energy loss of binary systems. The difference stems from the fact conservation violation equation in frequency domain gives rise to  a modified dispersion relation linking  the components of EM tensor with diffusion function. We assume that the diffusion function is also dependent on the reduced mass of the binary system with $\zeta$ being the dimensionless coupling constant. Working in the frequency domain, we derive closed form analytic expressions for the energy loss, explicitly depending on $\zeta$. 

The theory parameter $\zeta$ is constrained by computing the quantities  $\Delta_{\text{Obs}}$ and $\Delta_{\text{UG}}$ where they are defined in Eqs.(\ref{Delta-obs}) and (\ref{Delta-UG}). The predicted values for GR are obtained using the Peters-Mathews formula.  To be astrophysically viable, we assume  $\Delta_{\text{UG}}< \Delta_{\text{Obs}}$. The astrophysical parameters for the observation are done for six binary systems. We infer from Table-\ref{tab:constraint} that the strongest bound ($\vert\zeta\rvert \leq 5\times 10^{-4}$)  comes from PSR B1913+16 (Hulse-Taylor binary). Note that to compute the orbital period decay in UG, we assumed Newtonian gravity, which is a correct approximation since, in weak-gravity scenarios, UG reduces to Newtonian theory. This is because the polarization sum (\ref{eq:polarization-sum}) is identical to GR, which also reduces to Newtonian gravity in this limit. The obtained constrain on $\zeta$ is stronger by an order of magnitude than an equivalent constraint coming from the tidal deformability of the neutron stars \cite{Yang:2022ees} as we explained in the previous section.

In conclusion, let us specify some novelties of the present work. Due to the presence of a integration term which can be interpreted to give rise to the cosmological constant, most works in UG have focused on its cosmological consequences. However, our present work provides an alternative outlook by exploring its implications in astrophysical binaries. As of now, to the best of our knowledge, our obtained constrain on the non-conservation parameter of UG is the best astrophysical constraint. 

\section*{Acknowledgements}
Research of SJ is partially supported by the SERB, DST, Govt. of India, through a TARE fellowship grant no. TAR/2021/000354, hosted by the department of Physics, Indian Institute of Technology Kharagpur. I. C. thanks T. Parvez and S. Mandal for discussions.

\appendix

\section{Invariance of matter-graviton interaction term} \label{app:matter-graviton}

The interaction term for the matter-graviton vertex is: $-\dfrac{\kappa}{2}T_{\mu\nu} h^{\mu\nu}$.  We henceforth try to establish the gauge invariance of the interaction term in the restricted set of transverse diffeomorphisms for UG. The interaction vertex is given by,
\begin{equation}
    \mathcal{L}_{\text{int}}=-\frac{\kappa}{2} T_{\mu\nu} \,h^{\mu\nu}. 
\end{equation}
Upon a gauge transformation $x^\mu\rightarrow x^\mu+\xi^\mu$, where $\xi^\mu$ is given by Eq.(\ref{eq:xi_definition}), the graviton field transforms as,
\begin{equation}
    h_{\mu\nu} \rightarrow h_{\mu\nu} - 2 \partial_{(\mu} \, \xi_{\nu)}.
\end{equation}
The transformation in the interaction Lagrangian becomes,
\begin{align}
    T_{\mu\nu} \,h^{\mu\nu} \rightarrow
    & T_{\mu\nu} \,h^{\mu\nu} -  2\,  T_{\mu\nu}\, \,  \partial^{(\mu} \, \xi^{\nu)} \nonumber\\
     = & T_{\mu\nu} \,h^{\mu\nu} - 2\,  T_{\mu\nu}\, \, \partial^{(\mu} \epsilon^{\nu)\sigma\alpha\beta}\, \partial_{\sigma}\, \omega_{\alpha\beta} \nonumber\\
     = & T_{\mu\nu} \,h^{\mu\nu} - 2 \, \partial_{\sigma}\, \partial^{(\mu} T_{\mu\nu}\, \epsilon^{\nu)\sigma\alpha\beta}\, \omega_{\alpha\beta} \nonumber\\
     = & T_{\mu\nu} \,h^{\mu\nu} - 2 \,\partial_{\sigma}\, J^{(\mu} \,\epsilon^{\nu)\sigma\alpha\beta}\, \omega_{\alpha\beta}  \nonumber\\
     = & T_{\mu\nu} \,h^{\mu\nu} - 2 \bold{d} J \nonumber\\
     = & T_{\mu\nu} \,h^{\mu\nu}.
\end{align}
where $\partial^\mu \, T_{\mu\nu}= J_\nu$ and $ \bold{d} J =0 $ was established in Sec.\ref{sec:EM-violation}. 

This shows that even if the EM conservation does not hold in UG, the interaction Lagrangian remians invariant under transverse diffeomorphisms.

\bibliography{Final}
\end{document}